\documentclass[
 aps, 
 prb, 
 twocolumn,
 10pt,
 amsmath,
 amssymb,
 superscriptaddress,
 floatfix
]{revtex4-2}
\usepackage[T1]{fontenc}
\usepackage[utf8]{inputenc}
\usepackage{lmodern}
\usepackage{bm}
\usepackage{booktabs}
\usepackage{amsmath}
\usepackage{graphicx} 
\usepackage{enumitem}
\usepackage{multirow}
\usepackage{makecell}
\usepackage{siunitx}
\usepackage[
    colorlinks=true,
    citecolor=blue,
    urlcolor=blue
]{hyperref}
\providecommand\vect\bm

\begin{document}

\title{Inferring Magnetic Material Parameters from Statistical Measures in Strongly Fluctuating Magnetization Dynamics}

\author{Kübra Kalkan}
\thanks{These authors contributed equally to this work.}
\affiliation{Faculty of Physics, University of Duisburg-Essen, 47057 Duisburg, Germany}
\affiliation{Center for Nanointegration Duisburg-Essen (CENIDE), University of Duisburg-Essen, 47057 Duisburg, Germany}

\author{Atreya Majumdar}
\thanks{These authors contributed equally to this work.}
\affiliation{Faculty of Physics, University of Duisburg-Essen, 47057 Duisburg, Germany}
\affiliation{Center for Nanointegration Duisburg-Essen (CENIDE), University of Duisburg-Essen, 47057 Duisburg, Germany}

\author{Ross Knapman}
\affiliation{Faculty of Physics, University of Duisburg-Essen, 47057 Duisburg, Germany}
\affiliation{Center for Nanointegration Duisburg-Essen (CENIDE), University of Duisburg-Essen, 47057 Duisburg, Germany}
\affiliation{Institute of Mechanics, University of Duisburg-Essen, 45141 Essen, Germany}

\author{Omer Fetai}
\affiliation{Faculty of Physics, University of Duisburg-Essen, 47057 Duisburg, Germany}
\affiliation{Center for Nanointegration Duisburg-Essen (CENIDE), University of Duisburg-Essen, 47057 Duisburg, Germany}
\affiliation{Department of Economics, University of Bamberg, 96047, Bamberg, Germany.}

\author{Franziska Scheibel}
\affiliation{Functional Materials, Institute of Material Science, Technical University of Darmstadt, 64287 Darmstadt, Germany}

\author{Sabrina Disch}
\affiliation{Center for Nanointegration Duisburg-Essen (CENIDE), University of Duisburg-Essen, 47057 Duisburg, Germany}
\affiliation{Faculty of Chemistry, University of
Duisburg-Essen, 45141 Essen,
Germany}

\author{Illia Horenko}
\affiliation{Department of Mathematics, Artificial Intelligence in Mathematics, TU Kaiserslautern, Kaiserslautern, Germany}

\author{Karin Everschor-Sitte}
\affiliation{Faculty of Physics, University of Duisburg-Essen, 47057 Duisburg, Germany}
\affiliation{Center for Nanointegration Duisburg-Essen (CENIDE), University of Duisburg-Essen, 47057 Duisburg, Germany}

\begin{abstract}
Magnetic material parameters such as the exchange stiffness and magnetic anisotropy govern the behavior and functionality of magnetic systems, yet their local inference from magnetization data remains challenging, particularly in strongly fluctuating regimes with polycrystalline or multiphase microstructure, where conventional texture-based methods become unreliable. We introduce a magnetization-only framework for inferring material parameters from thermally driven magnetization dynamics. Using micromagnetic simulations, we extract statistical quantities such as temporal mean and latent entropy from the magnetization dynamics, fit models to these descriptors, and invert the models to infer material parameters. We show that this framework enables material-parameter inference as well as grain-boundary detection in a heterogeneous sample. Among the descriptors considered, latent entropy yields more accurate parameter estimates than the temporal mean. Our results establish latent entropy as an efficient descriptor for inferring magnetic material parameters from dynamical magnetization data and point toward its use for experimental parameter extraction at high temperatures and, more broadly, under strongly fluctuating conditions.

\end{abstract}

\maketitle

%%%%%%%%%%%%%%%%%%%%%%%%%%%%%%%%%%%%%%%%%%%%%%
%%%%%%%%%%%%%%%%%%%%%%%%%%%%%%%%%%%%%%%%%%%%%%
\section{Introduction}
%%%%%%%%%%%%%%%%%%%%%%%%%%%%%%%%%%%%%%%%%%%%%%
%%%%%%%%%%%%%%%%%%%%%%%%%%%%%%%%%%%%%%%%%%%%%%

\begin{figure*}[!tbp]
\includegraphics[width=1.0\linewidth]{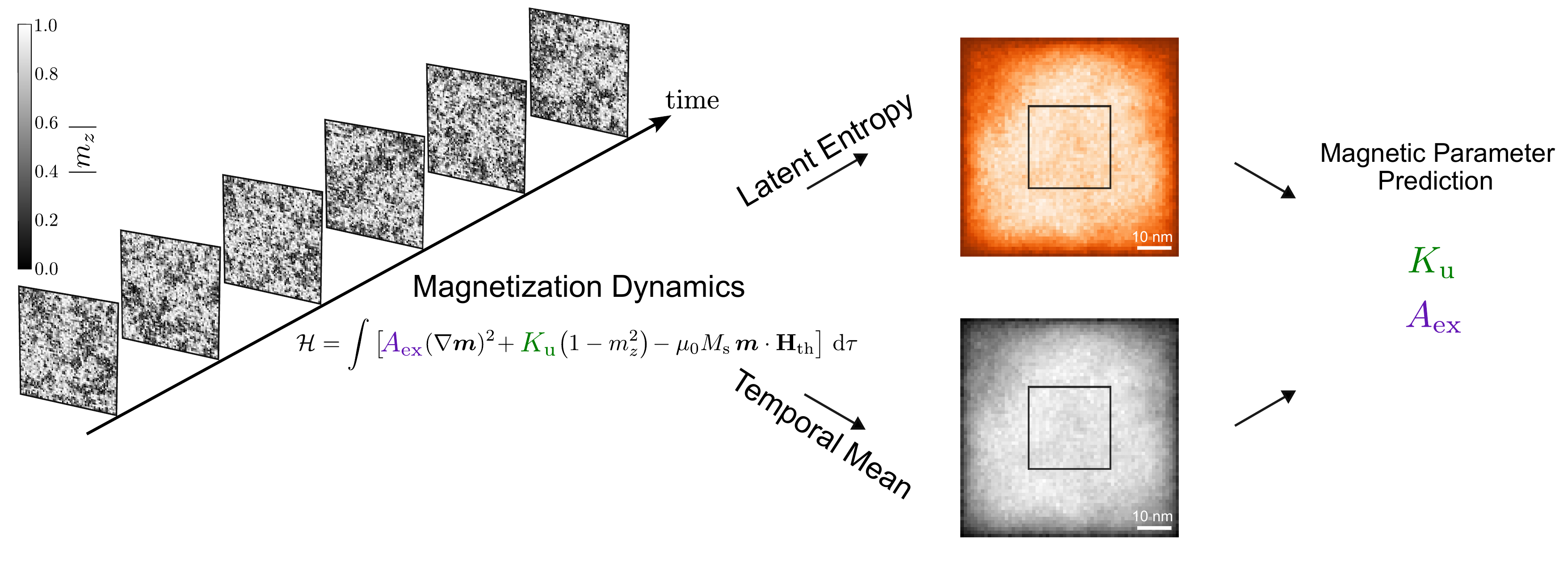} 
\caption{Workflow for inferring magnetic material parameters from thermally driven magnetization dynamics. Left: representative time-resolved snapshots of the out-of-plane magnetization magnitude $|m_z(\mathbf{r},t)|$ for a system with $A_{\rm ex}=8.2\times10^{-12}\,\mathrm{J\,m^{-1}}$, $K_{\rm u}=5.8\times10^{5}\,\mathrm{J\,m^{-3}}$, and $T=700\,\mathrm{K}$. Middle: pixel-resolved maps of the latent entropy $S$ and temporal mean $\mu$ computed from the magnetization time series; the outlined central region indicates the region of interest used to suppress edge effects. Right: the resulting descriptors are used within our framework to infer either $A_{\rm ex}$ for known $K_{\rm u}$ or $K_{\rm u}$ for known $A_{\rm ex}$.}
    \label{fig:Fig1}
\end{figure*}

Many functional properties of magnetic materials relevant to magnetic data storage, unconventional computing, and magnetocaloric applications are governed by intrinsic parameters such as the exchange stiffness and magnetic anisotropy~\cite{Bertotti1998_Ch5, coey2010magnetism, hirohata2020review, tishin2003the, gutfleisch2016mastering}. Quantitative knowledge of these parameters is therefore essential for understanding and controlling magnetic behavior at the local scale. 
However, determining these parameters \emph{locally} remains a major challenge. Conventional approaches infer them indirectly from magnetization textures, hysteresis, or magnetoresistance measurements, whereas highly localized techniques such as TEM-based off-axis electron holography require specialized specimen preparation that may alter the magnetic state~\cite{Bertotti1998_Ch5, slonczewski1973theory, ye2013determination, duninBorkowski2019}.
Inferring the material parameters from such indirect descriptors becomes particularly challenging in strongly fluctuating regimes, where thermal noise suppresses magnetization textures, rendering conventional approaches such as domain wall width extraction ineffective (Fig.~\ref{fig:Fig1}). In addition, domain walls often pin at defects or grain boundaries, which can alter their shape and limit their reliability as probes of the underlying material properties~\cite{yu1999pinning, hadjipanayis1988domain}. Taken together, these limitations hinder parameter extraction from imaging data in many experimentally relevant systems at elevated temperatures~\cite{niitsu2020temperature, kuch2011thermal}.

In this work, we introduce a magnetization dynamics-based framework that maps thermally driven magnetization dynamics to the underlying magnetic material parameters using statistical descriptors. Specifically, we infer either the exchange stiffness or the magnetocrystalline uniaxial anisotropy while treating the other parameter as known. Central to our approach is latent entropy, a computationally efficient, physics-based descriptor that quantifies the stochasticity of temporally ordered transitions between discretized magnetization states~\cite{Horenko2021, rodrigues2021deeper}. By incorporating information about how the system evolves between states, latent entropy captures dynamical features beyond simple amplitude-based measures and provides more informative signatures of the underlying exchange stiffness and anisotropy than the temporal mean. We study a parameter range chosen to cover values relevant for technologically important materials such as CoFe$_2$O$_4$ and L1$_0$-ordered FeNi~\cite{Eskandari2017, Suzuki1996, Frisk2017, Kovacs2020, Woodgate2023}. We apply the framework to spatially heterogeneous systems, enabling local parameter inference and the identification of grain boundaries directly from magnetization dynamics. These results demonstrate that reliable material-parameter inference is possible in strongly fluctuating regimes where conventional, texture-based approaches fail~\cite{kronseder2015real}. 

The remainder of the manuscript is organized as follows. In Section~\ref{sec:Methods}, we present the micromagnetic simulations and the calculation of the latent entropy. Section~\ref{sec:emp_model} details the empirical modeling of the descriptors based on the exchange and anisotropy values, and the scheme of predictive inference. Finally, Section~\ref{sec:inference_of_mat_parameters} presents our results for the prediction of the parameter values for uniform and heterogeneous samples.

%%%%%%%%%%%%%%%%%%%%%%%%%%%%%%%%%%%%%%%%%%%%%%
%%%%%%%%%%%%%%%%%%%%%%%%%%%%%%%%%%%%%%%%%%%%%%
\section{Methods}
\label{sec:Methods}
%%%%%%%%%%%%%%%%%%%%%%%%%%%%%%%%%%%%%%%%%%%%%%
%%%%%%%%%%%%%%%%%%%%%%%%%%%%%%%%%%%%%%%%%%%%%%

\subsection{Micromagnetic Simulations}
We simulate the magnetization dynamics $\boldsymbol{m}(\boldsymbol{r}, t)$ of a ferromagnetic thin film with perpendicular magnetic anisotropy whose micromagnetic model is given by the Hamiltonian
\begin{equation}
\mathcal H[\bm m]
= \int \!\Big[ A_{\rm ex}(\nabla\bm m)^2
+ K_{\rm u}\big(1-m_z^2\big)-\mu_0 M_{\rm s}\bm{m}\cdot\mathbf{H_{\rm th}} \Big]\, \mathrm{d}^3\boldsymbol{r} ,
\end{equation}
where $A_{\rm ex}$ and $K_{\rm u}$ represent the exchange stiffness and the uniaxial anisotropy constants, respectively. Simulations are performed using the GPU-accelerated micromagnetic solver \texttt{MuMax3}~\cite{Vansteenkiste2014}. The sample is initialized in a uniformly magnetized state along the out-of-plane $z$-direction. The thermal contribution to the dynamics is included via a stochastic, zero-mean thermal field $\mathbf{H}_{\mathrm{th}}$~\cite{Brown1963, Leliaert2017, Lyberatos1993}, whose variance is set by the fluctuation-dissipation theorem and therefore increases linearly with the absolute temperature $T$ (see Appendix~\ref{app:micromagnetic_simulations} for details). The temperature is set to $T=700~\mathrm{K}$, which remains below the Curie temperatures of CoFe$_2$O$_4$ and L1$_0$-ordered FeNi/tetrataenite~\cite{Eskandari2017, Woodgate2023},
while placing the system in a strongly fluctuating regime. The resulting thermal agitation also allows the magnetization trajectories to sample the full range of discretized states used in the latent entropy calculation, thereby avoiding empty sampled bins and enabling consistent estimation of the transition probabilities across all simulations. 

The magnetization evolves following the Landau-Lifshitz-Gilbert equation~\cite{Vansteenkiste2014, Gilbert2004} \begin{equation}
\frac{d \bm{m}}{dt}
= -\frac{\gamma}{1+\alpha^2} \bm{m}  \times\mathbf{H}_{\rm eff}
- \frac{\gamma\alpha}{1+\alpha^2}\,\bm{m} \times(\bm{m} \times \mathbf{H}_{\rm eff}).
\end{equation}
Here, $\gamma=\mu_0|\gamma_{\rm LL}|$ is the gyromagnetic constant
corresponding to an effective field expressed in $\mathrm{A\,m^{-1}}$,
$\alpha$ is the Gilbert damping parameter, and the effective field is given by $\mathbf{H}_{\rm eff} = -(\mu_0 M_s)^{-1} (\delta \mathcal{H} / \delta \bm{m})$,
where $\mu_0$ is the permeability of vacuum and $M_{\rm s}$ is the saturation magnetization.
The magnetization configuration is recorded every 1~ps over a simulation of duration 4.5~ns. For our analysis, the first $500$ frames $(500\,\mathrm{ps})$ are removed to eliminate the transient caused by the chosen initial state.
Because the magnetic easy axis is parallel to the out-of-plane direction of the film, we choose our observable to be the magnitude of the out-of-plane component, $|m_z(t)|$. Further micromagnetic simulation parameters are detailed in Appendix~\ref{app:micromagnetic_simulations}.

\begin{figure*}[!tbp]
    \includegraphics[width=1.0\linewidth]{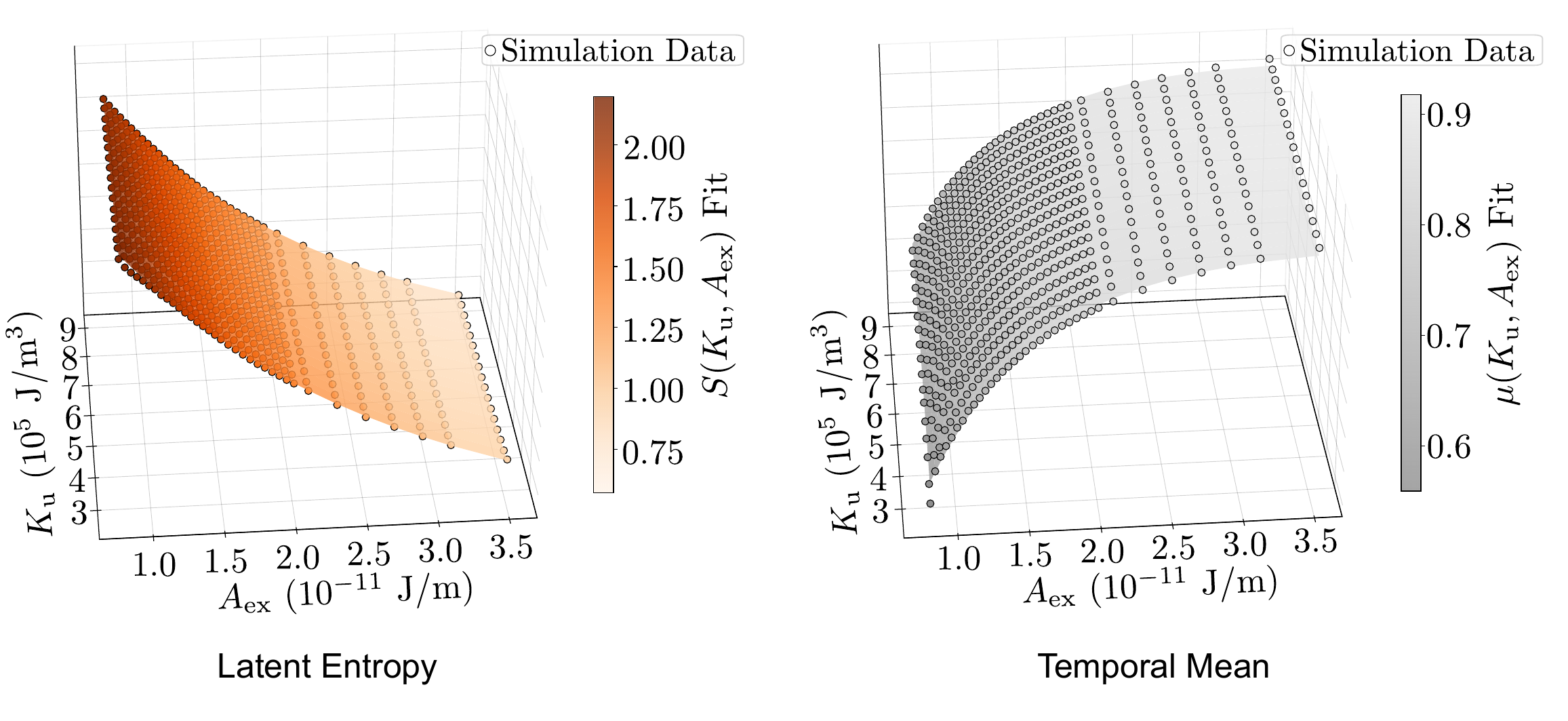} 
    \caption{Forward models relating the statistical descriptors to the magnetic material parameters. The left and right panels show the fitted latent entropy $S(K_{\rm u},A_{\rm ex})$ and temporal mean $\mu(K_{\rm u},A_{\rm ex})$, respectively, as functions of the exchange stiffness $A_{\rm ex}$ and uniaxial anisotropy $K_{\rm u}$. The surface plots represent the analytical fits given by Eq.~\eqref{eq:S-model} and~\eqref{eq:mu-model}, while the markers denote the values computed from the micromagnetic simulations.} 
    \label{fig:Fig2}
\end{figure*}

\subsection{Latent Entropy}
\label{latent_entropy}
We compute the pixel-wise latent entropy from the magnitude of the out-of-plane magnetization $|m_z(t)|$. The calculation begins by coarse-graining the continuous magnetization trajectory into a finite number ($N_{\rm bins}=9)$ of discrete states. This discretization allows the quantification of transition probabilities between the discrete states, which are analyzed within the latent process framework of~\cite{Horenko2021, rodrigues2021deeper}.
The maximum and minimum $|m_z|$-values used to define the bin edges are kept fixed across all simulations to ensure consistency in the latent entropy calculation. The latent entropy is then obtained from these state-to-state transition probabilities~\cite{Horenko2021}. 

Physically, a large latent entropy reflects highly stochastic, thermally agitated magnetization dynamics, whereas a small value indicates more 
predictable temporal evolution constrained by the energy landscape. Thus, the latent 
entropy serves as a sensitive descriptor of how the magnetization explores its accessible microstates, revealing signatures of the underlying material parameters such as the exchange stiffness $A_{\rm ex}$ and anisotropy $K_{\rm u}$. This measure captures dynamical features based on the transitions between discretized states and consequently, goes beyond standard statistics like the temporal mean.

%%%%%%%%%%%%%%%%%%%%%%%%%%%%%%%%%%%%%%%%%%%%%%
%%%%%%%%%%%%%%%%%%%%%%%%%%%%%%%%%%%%%%%%%%%%%%
\section{Inference Strategy}
\label{sec:emp_model}
%%%%%%%%%%%%%%%%%%%%%%%%%%%%%%%%%%%%%%%%%%%%%%
%%%%%%%%%%%%%%%%%%%%%%%%%%%%%%%%%%%%%%%%%%%%%%

To construct the reference dataset used to fit the empirical models, we simulate a $64~\mathrm{nm}\times64~\mathrm{nm}\times1~\mathrm{nm}$ sample in which the exchange stiffness and uniaxial anisotropy are spatially uniform. From the resulting magnetization dynamics, we compute the pixel-wise temporal mean, $\mu=\langle |m_z(t)| \rangle$, and latent entropy, $S(|m_z(t)|)$. These quantities are then averaged over a central square region defined by excluding a $20$~nm-wide margin from each edge, as shown in Fig.~\ref{fig:Fig1}. The excluded margin exceeds the longest exchange length ($\sim \sqrt{A^{max}_{\rm ex}/K^{min}_u}=11.6~\rm{nm}$) considered in our study and thereby suppresses boundary artifacts (Table.~\ref{tab:sim_parameters}). Each simulation therefore yields two scalar descriptors, $\mu$ and $S$, for the selected material-parameter pair. We repeat this procedure while systematically varying the exchange stiffness and anisotropy over the parameter ranges $[8.2\times10^{-12}, 35\times10^{-12}]\; \mathrm{J}/\mathrm{m}$ and $[2.6\times10^5, 9\times 10^5]\; \mathrm{J}/\mathrm{m}^3$, respectively, generating a dataset used for the subsequent modeling. 
Following this, we use two empirical forward models to fit the latent entropy and mean as a function of the exchange stiffness and uniaxial anisotropy (Fig.~\ref{fig:Fig2}) as:
\begin{align} \label{eq:S-model}
    S(K_{\rm u}, A_{\rm ex}) &=S_0 
     +  S_1 \cdot \exp\!\left[-\left(S_2 + S_3 K_{\rm u}^{S_4}\right) A_{\rm ex}^{S_5}\right]
\end{align}
and 
\begin{equation}  \label{eq:mu-model}
    \begin{aligned}
    \mu(K_{\rm u}, A_{\rm ex}) = \;&
    \mu_0 - \mu_1 \exp\!\left(-\mu_2 A_{\rm ex}\right) \\
    & + \mu_3 A_{\rm ex}^{-\mu_4}
    \left[1 - \exp\!\left(-\mu_5 K_{\rm u}^{\mu_6}\right)\right]).
    \end{aligned}
\end{equation}
The fitted forward models capture the simulation data with high accuracy, yielding coefficients of determination of $R^2=0.9997$ for the latent entropy model and $R^2=0.9962$ for the temporal mean model. Details on the fit parameters and fit metrics are provided in Appendix~\ref{app:fitting_details}.

To infer the material parameters, we assume one parameter to be known and reconstruct the other by inverting the corresponding fitted model. For a sample with observed latent entropy $S_{\mathrm{obs}}$, the exchange stiffness and uniaxial anisotropy are obtained as
\begin{subequations}
\begin{align} \hat{A}_{\rm ex} &= S^{-1}(S_{\mathrm{obs}};K_{\rm u}),\\
\hat{K}_u &= S^{-1}(S_{\mathrm{obs}};A_{\rm ex}). \label{eq:parameter_inv} 
\end{align}
\end{subequations}
respectively. The $S$-model admits closed-form inverses for both $A_{\rm ex}$ and $K_{\rm u}$, obtained by algebraic rearrangement of Eq.~\eqref{eq:S-model}. For the $\mu$-model, $K_{\rm u}$ can likewise be obtained analytically, whereas $A_{\rm ex}$ is determined numerically by solving $\mu(K_{\rm u},A_{\rm ex})=\mu_{\mathrm{obs}}$ using Brent's method~\cite{brent1973algorithms}, as implemented in \texttt{scipy.optimize.root\_scalar}~\cite{virtanen2020scipy}. The numerical search for $A_{\rm ex}$ was bounded to the range $[2\times10^{-12}, 60\times10^{-12}]~\mathrm{J}/\mathrm{m}$.

%%%%%%%%%%%%%%%%%%%%%%%%%%%%%%%%%%%%%%%%%%%%%%
%%%%%%%%%%%%%%%%%%%%%%%%%%%%%%%%%%%%%%%%%%%%%%
\section{Inference of material parameters}
\label{sec:inference_of_mat_parameters}
%%%%%%%%%%%%%%%%%%%%%%%%%%%%%%%%%%%%%%%%%%%%%%
%%%%%%%%%%%%%%%%%%%%%%%%%%%%%%%%%%%%%%%%%%%%%%

In this section, we demonstrate how statistical descriptors extracted from magnetization dynamics can be used to infer material parameters, first for spatially uniform samples and subsequently for spatially heterogeneous systems containing grain boundaries.

\begin{figure*}[htbp]
\includegraphics[width=0.8\linewidth]{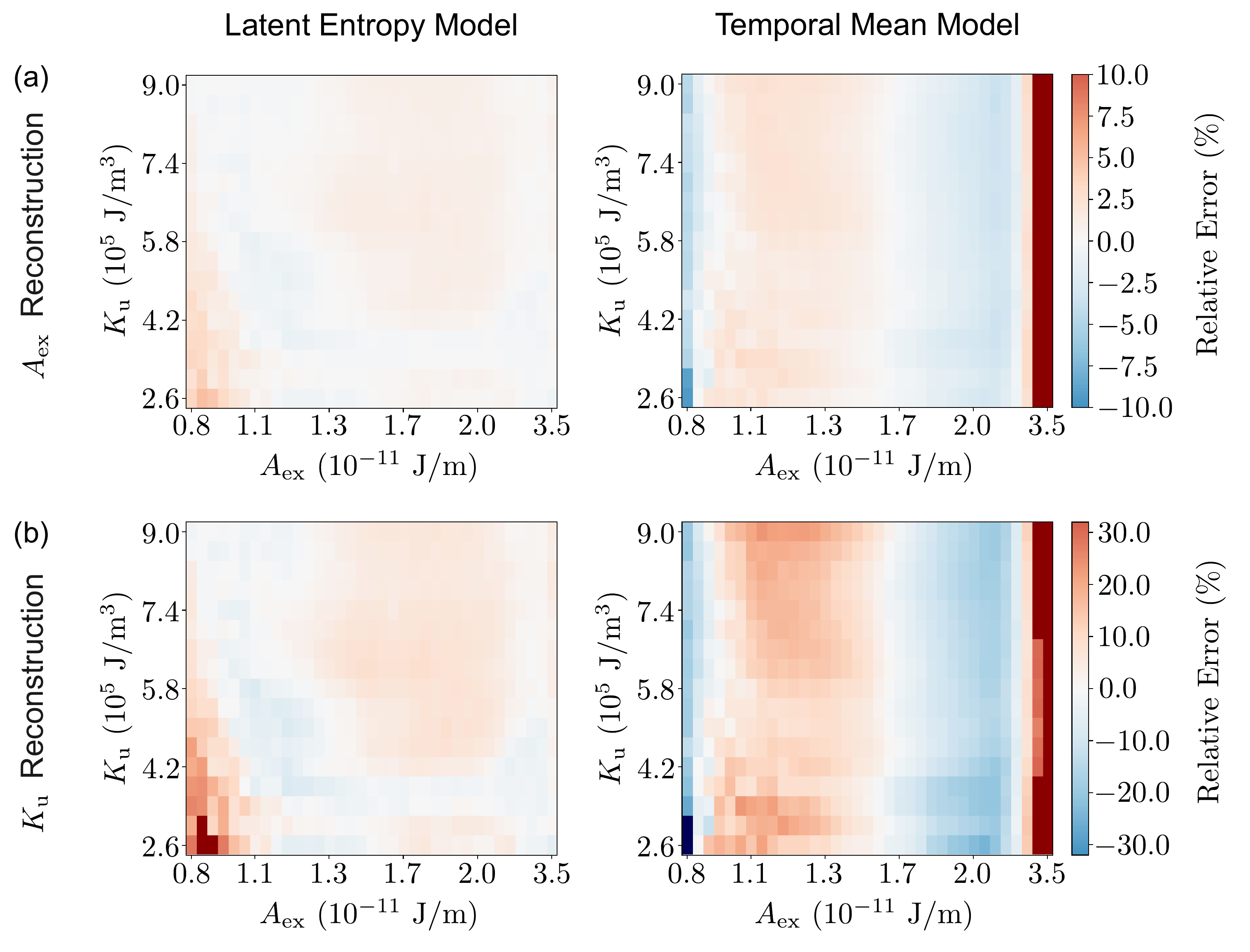} 
\caption{Relative reconstruction errors across the uniform-sample parameter grid. The error is defined as $(\hat{p}-p)/p\times100\%$, where $p$ denotes the true parameter value and $\hat{p}$ its reconstructed value. The left and right columns show results obtained from the $S$- and $\mu$-models, respectively. (a) Reconstruction of $A_{\rm ex}$ with $K_{\rm u}$ treated as known. (b) Reconstruction of $K_{\rm u}$ with $A_{\rm ex}$ treated as known. The color scales are limited to $\pm10\%$ in (a) and $\pm32\%$ in (b); values outside these ranges are clipped to the darkest colors.}
   \label{fig:Fig3}
\end{figure*}

\subsection{Reconstruction in Uniform Samples}
\label{subsec:iv_a}
To assess the inference accuracy in spatially uniform systems, we reconstruct $A_{\rm ex}$ and $K_{\rm u}$ values across the full parameter range using the inversion scheme introduced in Section~\ref{sec:emp_model}. The resulting relative error maps are shown in Fig.~\ref{fig:Fig3}, with panel (a) demonstrating that the latent entropy model reconstructs $A_{\rm ex}$ more accurately than the temporal mean model, yielding average absolute relative errors of $0.66\%$ and $4.79\%$, respectively. Except in the low-$A_{\rm ex}$, low-$K_{\rm u}$ regime, the errors of the latent entropy model remain within approximately $\pm 2\%$ as shown in Appendix~\ref{app:parameterinference_reference}. The largest relative reconstruction error, approximately $5.0\%$, occurs in the low-$A_{\rm ex}$, low-$K_{\rm u}$ regime, where strong thermal fluctuations reduce the dynamical contrast between neighboring parameter combinations.
By contrast, the errors of the temporal mean model are more broadly distributed: although most lie within approximately $\pm5\%$, a pronounced high-error region with absolute errors exceeding $10\%$ emerges at the largest $A_{\rm ex}$ values, shown in dark red in Fig.~\ref{fig:Fig3}(a). The corresponding error distribution is shown in Fig.~\ref{fig:FigC1}(a), right. A detailed statistical analysis of relative errors is described in Appendix~\ref{app:parameterinference_reference}.

\begin{figure*}[!tbp]
    \includegraphics[width=1.0\linewidth]{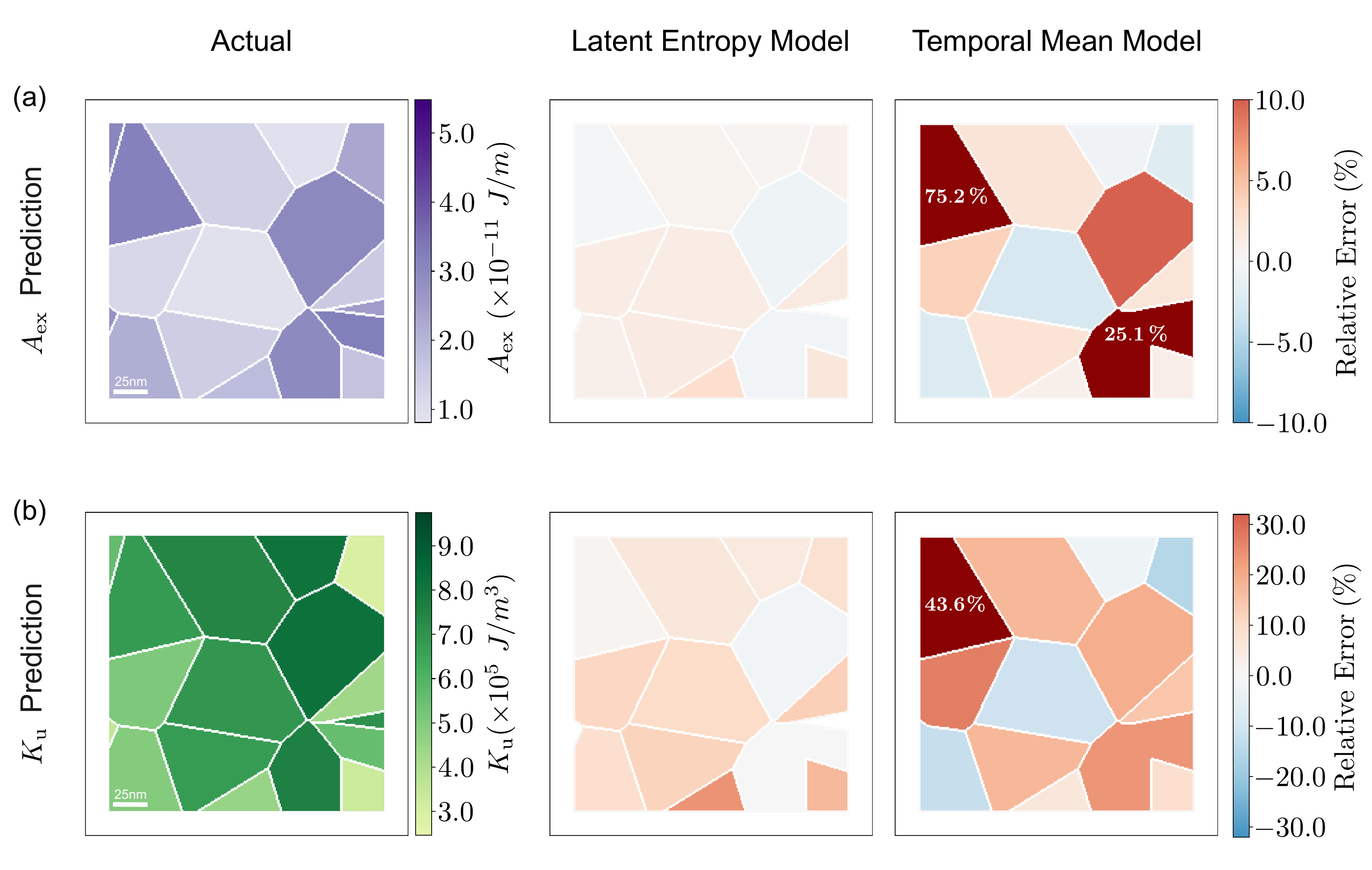} 
    \caption{Prediction of magnetic material parameters in the Voronoi-tessellated sample with each region having a uniform, randomly chosen ($A_{\rm ex}, K_{\rm u}$) pair. (a) Prediction of the exchange stiffness $A_{\rm ex}$ with $K_{\rm u}$ treated as known. (b) Prediction of the uniaxial anisotropy $K_{\rm u}$ with $A_{\rm ex}$ treated as known. The left column shows the imposed parameters in the different Voronoi regions, while the middle and right columns show the relative prediction errors obtained from the $S$- and $\mu$-models, respectively. As detailed in Appendix~\ref{app:Voronoi_geometry}, for each region, the predicted values are averaged over the grain-interior pixels and displayed across the full spatial extent of the region. The error scales are limited to $\pm10\%$ in (a) and $\pm32\%$ in (b); values outside these ranges are clipped to the darkest colors, and the values indicated explicitly. The white lines delineating the region boundaries are included solely as visual guides and do not indicate zero relative error.} 
    \label{fig:Fig4}
\end{figure*}

The reconstruction of $K_{\rm u}$, shown in Fig.~\ref{fig:Fig3}(b), exhibits a similar overall trend, with the $S$-model outperforming the $\mu$-model across most of the parameter range. However, the magnitudes of relative errors are larger than in the $A_{\rm ex}$ inference. The average absolute relative errors are $4.32\%$ for the $S$-model and $20.48\%$ for the $\mu$-model. For the $S$-based model, the relative errors are typically within $\pm 10\%$, with a maximum of $43.75\%$ occurring again in the low-$A_{\rm ex}$, low-$K_{\rm u}$ region (Fig.~\ref{fig:Fig3}(b), left). By contrast, the $\mu$-model presents a broader error distribution, where most errors lie within $\pm 20\%$ and absolute errors exceeding $32\%$ at the largest $A_{\rm ex}$ values (Fig.~\ref{fig:Fig3}(b), right).

The deterioration of the $\mu$-model reconstruction at high $A_{\rm ex}$ arises because $|m_z|$ approaches saturation in this regime. Strong exchange coupling suppresses the fluctuations in the magnetization magnitude, causing the temporal mean to vary only weakly with either material parameter and making the numerical inversion ill-conditioned. The latent entropy remains more informative because it captures the transition structure of the dynamics, which retains parameter-dependent information even when the mean magnetization changes only weakly.

The observed prediction errors are determined jointly by the residuals of the fitted forward models and their local sensitivities. 
As shown in App.~\ref{app:fitting_details}, the latent entropy fit exhibits comparatively small, spatially mixed residuals and
larger inference errors arise from weaker sensitivity to $K_{\rm u}$ and, for the $\mu$ model, from reduced sensitivity at high $A_{\rm ex}$.

\subsection{Grain Boundary Detection and Parameter Prediction in Spatially Heterogeneous Systems}
\label{subsec:iv-b}
We next examine whether the proposed framework can resolve spatially heterogeneous magnetic properties. To this end, we consider a sample of dimensions $256~{\rm nm} \times 256~{\rm nm} \times 1~{\rm nm}$, divide it into distinct Voronoi regions with an average size of approximately $2970~{\rm nm^2}$, and assign each region a spatially uniform but randomly selected pair $(K_{\rm u}, A_{\rm ex})$ from our parameter range (see Fig.~\ref{fig:Fig4}).
From the resulting magnetization dynamics, we compute pixel-resolved maps of the latent entropy $S$ and temporal mean $\mu$. The heterogeneous sample allows us to assess two complementary capabilities of the framework: the identification of grain boundaries (interfaces between regions) and the reconstruction of the material parameters within their interiors.

\paragraph{Grain boundary detection:}
Before performing local parameter prediction, we identify the interfaces between neighboring Voronoi grains from the spatial maps of the latent entropy and temporal mean. Because both descriptors are approximately uniform within individual grains but change across their interfaces, their spatial gradients provide a direct means of locating the boundaries. The detected boundary signals are converted into masks, and the remaining unmasked pixels define the grain interiors used for the subsequent parameter prediction. The procedure is described and illustrated in detail in Appendix~\ref{app:Voronoi_geometry}.

Overall, this gradient-based procedure reliably separates grain interiors from interfacial regions for both observables.
Differences in parameter sensitivity and spatial noise lead to observable-dependent masking performance. In particular, the $\mu$-based mask does not resolve some interfaces that are detected by the $S$-based mask and produces more irregular boundary contours, consistent with the noisier spatial gradients of the temporal-mean map.

\paragraph{Parameter prediction within regions:}
Following grain boundary localization, the remaining grain interiors are segmented into distinct regions and the inference scheme is implemented individually for each unmasked pixel. Before inversion, the detected boundary centerlines are expanded into 19~nm-wide exclusion regions, as described in Appendix~\ref{app:Voronoi_geometry}. This removes pixels near the grain boundaries, where the local parameter inference becomes unreliable because the magnetization dynamics are influenced by exchange coupling across neighboring grains.

As in the uniform samples, each Voronoi grain has uniform parameters; we average the pixel-wise predictions over its unmasked interior to obtain one representative value per grain.
 Fig.~\ref{fig:Fig4} compares the actual parameter values with the relative prediction errors obtained from the $S$- and $\mu$-models. The region-averaged predictions are extended across the full spatial extent of each grain, including the boundary zones excluded from the averaging, where the material parameters are assumed to remain unchanged.

The prediction results show that both descriptors can identify the grain boundaries and recover the spatial variation of the material parameters in the heterogeneous sample. Consistent with the trends observed for the spatially uniform samples, the $S$-model yields smaller errors than the $\mu$-model for both material parameters, and $A_{\rm ex}$ is predicted more accurately than $K_{\rm u}$. A detailed statistical analysis of the prediction accuracy across the different regions is presented in Appendix~\ref{app:Voronoi_geometry}.

%%%%%%%%%%%%%%%%%%%%%%%%%%%%%%%%%%%%%%%%%%%%%%
%%%%%%%%%%%%%%%%%%%%%%%%%%%%%%%%%%%%%%%%%%%%%%
\section{Discussion and Outlook}
%%%%%%%%%%%%%%%%%%%%%%%%%%%%%%%%%%%%%%%%%%%%%%
%%%%%%%%%%%%%%%%%%%%%%%%%%%%%%%%%%%%%%%%%%%%%%

The results show that thermally driven magnetization dynamics offer a practical approach to inferring intrinsic magnetic parameters, with latent entropy serving as a more informative descriptor than the temporal mean, particularly in strongly fluctuating regimes where conventional texture-based methods fail. This makes our approach particularly attractive for elevated-temperature measurements and industrially relevant operating conditions. 
The superior performance of the entropy-based inference strategy is attributed to the greater sensitivity of latent entropy to variations in the material parameters across a larger portion of the parameter space. At the same time, the temporal mean remains a useful complementary descriptor, as it captures the average magnitude of the magnetization fluctuations, whereas the latent entropy reflects their stochastic transition structure. Combining both quantities in a joint inference framework may therefore further improve predictive performance.

The extension to Voronoi-tessellated samples shows that the framework can identify grain boundaries directly from magnetization dynamics and reconstruct local material parameters within grain interiors. This capability is particularly relevant to magnetic materials, since grain boundaries are key microstructural features that strongly influence magnetic coupling, domain wall pinning, and magnetization reversal in real materials. 
Therefore, the models are not expected to perform reliably near interfaces, where boundary effects and coupling to neighboring grains violate the assumption that the local dynamics can be described by a single parameter set.
The masking procedure also imposes a practical spatial-resolution limit, since grains with characteristic dimensions comparable to or smaller than the 19~nm-wide exclusion regions may be fully masked and therefore cannot be resolved as independent regions.

The present results also point to a clear experimental perspective: the framework requires only time-resolved magnetization data and does not rely on well-formed domain walls, other easily identifiable textures, or hysteresis measurements. Experimentally, the required magnetization dynamics could be probed using techniques such as time-resolved magneto-optical Kerr microscopy, ferromagnetic resonance, Brillouin light scattering, time-resolved scanning transmission X-ray microscopy, and time-resolved XMCD-PEEM~\cite{hollander2017component, puzic2005spatially, sebastian2015micro, feggeler2023scanning, bolte2008time}. The computationally inexpensive nature of the temporal mean and latent entropy further supports the application of the method to large datasets and spatially resolved parameter mapping.

In three-dimensional systems, texture-based parameter extraction is especially challenging because the observed domain-wall width depends on both the local wall orientation and the projection used for imaging. Our systematic approach to parameter inference from highly fluctuating data can be extended naturally to three-dimensional magnetic datasets. By relying on local magnetization dynamics rather than on a specific projected domain-wall profile, the descriptor-based framework may therefore be particularly well suited to volumetric magnetic imaging.

Furthermore, the framework may also be relevant to more complex microstructures in which local inhomogeneities strongly affect the dynamics. Besides grain boundaries, another example is the presence of precipitates or secondary phases, which act as pinning centers and modify local reversal processes~\cite{polin2026direct}. Related interface-driven effects arise in magnetic nanoparticles and nanoclusters, where spin disorder at particle surfaces and interfaces can strongly influence magnetization reversal and magnetic heating efficiency~\cite{lak2021embracing}. In such systems, the magnetization dynamics can carry signatures not only of intrinsic parameters within a grain but also of local microstructural features that perturb the effective energy landscape. Extending the present approach to such settings could therefore provide a route toward identifying and characterizing microstructural heterogeneities from dynamical magnetic data.

Another natural direction is to move beyond purely empirical fitting functions. In the present work, the fitted forward models based on latent entropy and temporal mean are sufficient to demonstrate the inference concept. However, more physics-informed forward models could improve extrapolatability and interpretability, especially if the framework is extended to broader parameter ranges, additional interactions, or different material classes. 
For example, one could seek fitting forms motivated more directly by thermal activation, characteristic energy scales, or established micromagnetic scaling relations. Such models may also help clarify how the descriptors depend on exchange, anisotropy, and temperature in a more transparent way.

The present study focuses on thermally driven magnetization fluctuations in systems governed by exchange and anisotropy energy terms. Incorporating additional contributions, such as dipolar or Dzyaloshinskii-Moriya interactions, would therefore extend the framework's applicability to a broader class of magnetic systems. The present inference framework could also be combined with machine-learning-based segmentation approaches, such as U-Net models that localize defects from spatial maps of dynamical descriptors~\cite{knapman2026defect}. Such a combined workflow would first identify and segment material inhomogeneities and then apply the parameter-inference scheme within the detected regions to reconstruct their local magnetic properties.

Overall, the present results establish latent entropy-based inference as a promising route for extracting magnetic material parameters in regimes where conventional methods face limitations. The method is computationally cheap, naturally compatible with spatially resolved analysis, and readily extendable to richer descriptors and more physics-based forward models. 
These features establish a promising foundation for quantitative parameter mapping in realistic, noisy, and heterogeneous magnetic systems, ultimately enabling the knowledge-driven design and optimization of next-generation high-performance magnets.

%%%%%%%%%%%%%%%%%%%%%%%%%%%%%%%%%%%%%%%%%%%%%%
%%%%%%%%%%%%%%%%%%%%%%%%%%%%%%%%%%%%%%%%%%%%%%
\section{Acknowledgments}
%%%%%%%%%%%%%%%%%%%%%%%%%%%%%%%%%%%%%%%%%%%%%%
%%%%%%%%%%%%%%%%%%%%%%%%%%%%%%%%%%%%%%%%%%%%%%

We thank Nasim Bazazzadeh, Robin Msiska, Janine Graser, Maria Azhar, Alessandro Pignedoli, Yangyiwei Yang, and Tatiana Smoliarova for fruitful discussions. We acknowledge funding from the Deutsche Forschungsgemeinschaft (DFG, German Research Foundation) – Project‐ID 405553726 – CRC/TRR 270, projects B12, A12 and B11. A.~M.\ acknowledges support from the Joachim Herz Foundation through the Add-on Fellowship for Interdisciplinary Science and Transfer.

\appendix
\makeatletter
\@addtoreset{figure}{section}
\@addtoreset{table}{section}
\makeatother

\renewcommand{\thefigure}{\thesection\arabic{figure}}
\renewcommand{\thetable}{\thesection\arabic{table}}

\section{Micromagnetic Simulations}
\label{app:micromagnetic_simulations}
The thermal field $\mathbf{H}_{\mathrm{th}}$ is constructed in 
accordance with the fluctuation--dissipation theorem to ensure 
thermodynamically consistent finite-temperature dynamics.
\begin{equation}
\big\langle \mathbf{H}_{\mathrm{th}}(t) \big\rangle = \mathbf{0},
\end{equation}
and the Cartesian components $H_{\mathrm{th},i}$ 
(with $i,j \in \{x,y,z\}$) form a Gaussian white-noise process that is 
uncorrelated across directions and delta-correlated in time. The covariance is given by
\begin{equation}
\big\langle H_{\mathrm{th},i}(t)\, H_{\mathrm{th},j}(t') \big\rangle
=\frac{2 \alpha k_\mathrm{B} T}{\gamma_0 \mu_0 M_\mathrm{s} V\, }\,
\delta_{ij}\, \delta(t - t'),
\end{equation}
where $k_\mathrm{B}$ is the Boltzmann constant,
$V$ is the volume of a discretized simulation cell and $\gamma_0 = \SI{2.21e5}{\meter\per\ampere\per\second}$.

\begin{table}[h!]
    \centering
    \caption{Input Parameters for Micromagnetic Simulations}
    \setlength{\tabcolsep}{6pt}
    \renewcommand{\arraystretch}{1.2}
    \begin{tabular}{p{0.58\linewidth}p{0.33\linewidth}}%{c|c}
    \toprule
        Grid size (for reference dataset) & (64$\times$64$\times$1)\\
        Grid size (for heterogeneous prediction) & (256$\times$256$\times$1)\\
        Cell size & ($1\,\mathrm{nm}$, $1\,\mathrm{nm}$, $1\,\mathrm{nm}$)\\
        Temperature & 700~K\\
        $M_\mathrm{s}$ & 4$\times10^5$~A/m\\
            $K_{\rm u}$ range ($\times10^5$) & $2.6 \le K_{\rm u} \le 9\; \mathrm{J}/\mathrm{m}^3$\\
        $A_{\rm ex}$ range ($\times10^{-12}$) & $8.2 \le A_{\rm ex} \le 35\; \mathrm{J}/\mathrm{m}$\\
        Total runtime & $4.5\, \mathrm{ns}$\\
        Sampling interval & $1\, \mathrm{ps}$ \\
        $\alpha$ & 0.1\\
    \bottomrule
    \end{tabular} 
    \label{tab:sim_parameters}
\end{table}

\section{Empirical Fits and Sensitivity Analysis}
\label{app:fitting_details}
\subsection{Fitted Model Parameters}

The fit parameters entering Eqns.~\eqref{eq:S-model} and~\eqref{eq:mu-model} are listed in Table~\ref{tab:fit_parameters}. The parameter values are expressed in terms of the dimensionless rescaled variables
$A^*_{\rm ex} = A_{\rm ex}/(10^{-12}$~J/m) and $K^*_{\rm u} = K_{\rm u}/(10^5$~J/m$^3$).

\begin{table}[tbp]
\centering
\caption{Fit parameters for the latent entropy and temporal mean models.}
\label{tab:fit_parameters}
\small
\begin{minipage}[t]{0.46\columnwidth}
\centering
\textbf{Latent entropy model}\\[2pt]
$S(K_{\rm u},A_{\rm ex})$\\[4pt]
\begin{tabular}{@{}c r@{}}
\hline
Parameter & Value \\
\hline
$S_0$ & 0.34  \\
$S_1$ & 2.86  \\
$S_2$ & 0.366 \\
$S_3$ & 0.102 \\
$S_4$ & 0.40  \\
$S_5$ & 0.90  \\
\hline
\end{tabular}
\end{minipage}
\hfill
\begin{minipage}[t]{0.46\columnwidth}
\centering
\textbf{Temporal mean model}\\[2pt]
$\mu(K_{\rm u},A_{\rm ex})$\\[4pt]
\begin{tabular}{@{}c r@{}}
\hline
Parameter & Value \\
\hline
$\mu_0$ & 0.890 \\
$\mu_1$ & 1.01  \\
$\mu_2$ & 1.31  \\
$\mu_3$ & 0.113 \\
$\mu_4$ & 0.988 \\
$\mu_5$ & 0.125 \\
$\mu_6$ & 1.24  \\
\hline
\end{tabular}
\end{minipage}
\end{table}

\begin{figure*}[htbp]
    \includegraphics[width=0.8\linewidth]{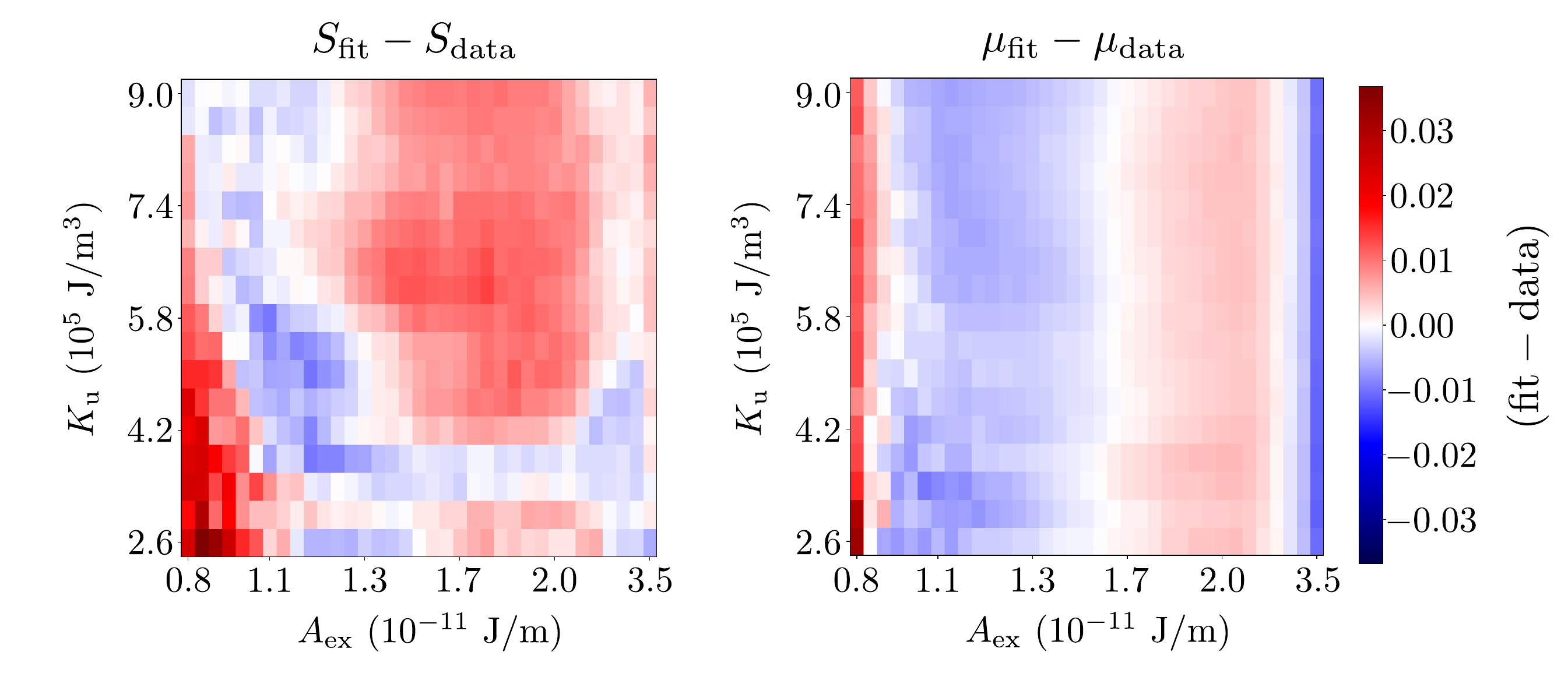} 
    \caption{Residual maps of the fitted forward models across the uniform-sample parameter grid. The left and right panels show the residuals $S_{\mathrm{fit}}-S_{\mathrm{data}}$ and $\mu_{\mathrm{fit}}-\mu_{\mathrm{data}}$, respectively, for the models defined in Equations~\eqref{eq:S-model} and~\eqref{eq:mu-model}.}
    \label{fig:FigB1}
\end{figure*}

The heatmap of residuals (fit~--~data) are shown in Fig.~\ref{fig:FigB1}. The quality of the fitted models is assessed using the coefficient of determination ($R^2$), root mean squared error (RMSE), mean absolute error (MAE), and maximum absolute relative error, as summarized in Table~\ref{tab:entropy_mean_model_metrics}.

\begin{table}[ht]
\centering
\caption{Global fit quality metrics for the forward models: latent entropy $S(K_{\rm u}, A_{\rm ex})$ and temporal mean $\mu(K_{\rm u}, A_{\rm ex})$.}
\begin{tabular}{l c c}
\hline
Metric & $S$-model & $\mu$-model \\
\hline
$R^2$ & $0.9997$ & $0.9962$ \\
RMSE   & $6.36 \times 10^{-3}$ & $4.81 \times 10^{-3}$  \\
MAE    & $4.98 \times 10^{-3}$ & $3.79 \times 10^{-3}$ \\
$\rm max|error|$ & $0.0341$ & $0.0320$ \\
$\rm mean|rel \, error|$ & $3.13 \times 10^{-3}$ & $5.00 \times 10^{-3}$ \\
$\rm max|rel \,error|$ & $1.62 \times 10^{-2}$ & $5.72 \times 10^{-2}$ \\

\hline
\end{tabular}
\label{tab:entropy_mean_model_metrics}
\end{table}

\subsection{Sensitivity Analysis}
\label{app:appB_sensitivityanalysis}

\begin{figure*}[htbp]
    \includegraphics[width=0.6\linewidth]{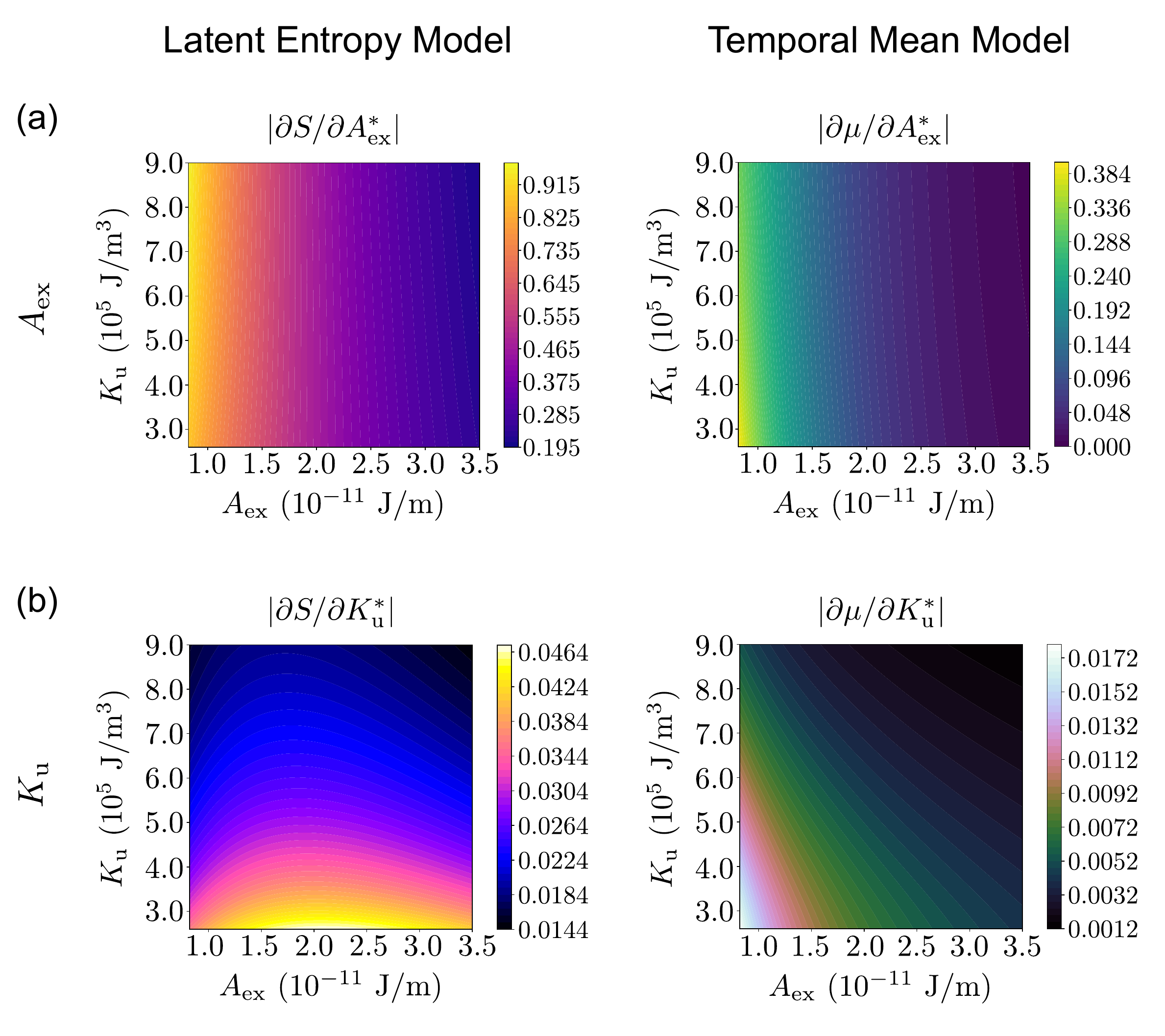} 
    \caption{Sensitivity analysis of the latent entropy and temporal mean forward models across the reference parameter space. The top row shows the absolute derivatives with respect to $A^*_{\rm ex}$, and the bottom row shows the corresponding derivatives with respect to $K^*_{\rm u}$, for the $S$-model (left) and $\mu$-model (right).}
    \label{fig:FigB2}
\end{figure*}

To quantify how strongly the fitted observables respond to variations in the material parameters, we evaluate the sensitivities: the absolute partial derivatives of the forward models with respect to the dimensionless variables $A_{\rm ex}^*$ and $K_{\rm u}^*$ across the investigated parameter space. Fig.~\ref{fig:FigB2} presents the sensitivity of $S$ and $\mu$ to exchange and anisotropy. The sensitivity maps show that both observables respond more strongly to variations in $A_{\rm ex}$ than in $K_{\rm u}$, while the sensitivity of $\mu$-model to $K_{\rm u}$ decreases with increasing $A_{\rm ex}$.

\section{Distribution of Reconstruction Errors in Uniform Samples}
\label{app:parameterinference_reference}

\begin{figure*}[htbp]
    \includegraphics[width=0.9\linewidth]{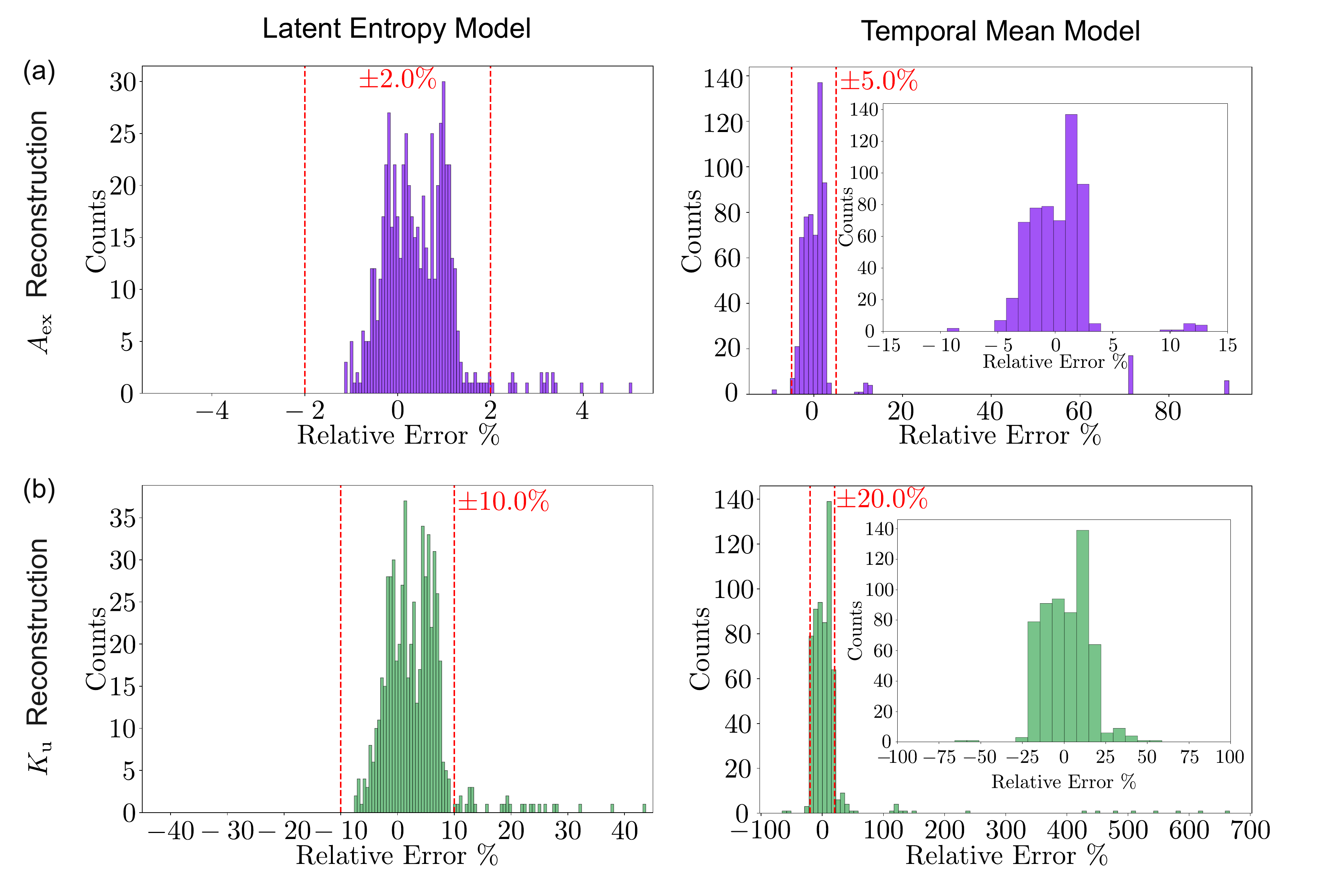}
    \caption{Distributions of the relative reconstruction errors across the uniform-sample parameter grid shown in Figure~\ref{fig:Fig3}. The top and bottom rows correspond to the reconstruction of $A_{\rm ex}$ and $K_{\rm u}$, respectively, while the left and right columns show results from the $S$- and $\mu$-models. The dashed vertical lines indicate intervals containing approximately $90\%$ of the reconstructed values: $\pm2\%$ and $\pm5\%$ for $A_{\rm ex}$, and $\pm10\%$ and $\pm20\%$ for $K_{\rm u}$, from the $S$- and $\mu$-models, respectively. Insets in the $\mu$-model panels show magnified views of the central parts of the distributions.}
    \label{fig:FigC1}
\end{figure*}

Fig.~\ref{fig:FigC1} shows that the $S$-model yields narrower relative error distributions than the $\mu$-model for both inferred parameters. For $A_{\rm ex}$, approximately $90\%$ of the errors lie within $\pm2\%$ for the $S$-model and within $\pm5\%$ for the $\mu$-model. The $K_{\rm u}$ distributions are broader, with corresponding intervals of approximately $\pm10\%$ and $\pm20\%$, reflecting the lower sensitivity of both observables to anisotropy. The $\mu$-model additionally exhibits pronounced tails for both parameter reconstructions.

\section{Masking procedure and Distribution of Prediction Errors}
\label{app:Voronoi_geometry}
\subsection{Boundary Masking Procedure}
\begin{figure*}[htbp]
    \includegraphics[width=1.0\linewidth]{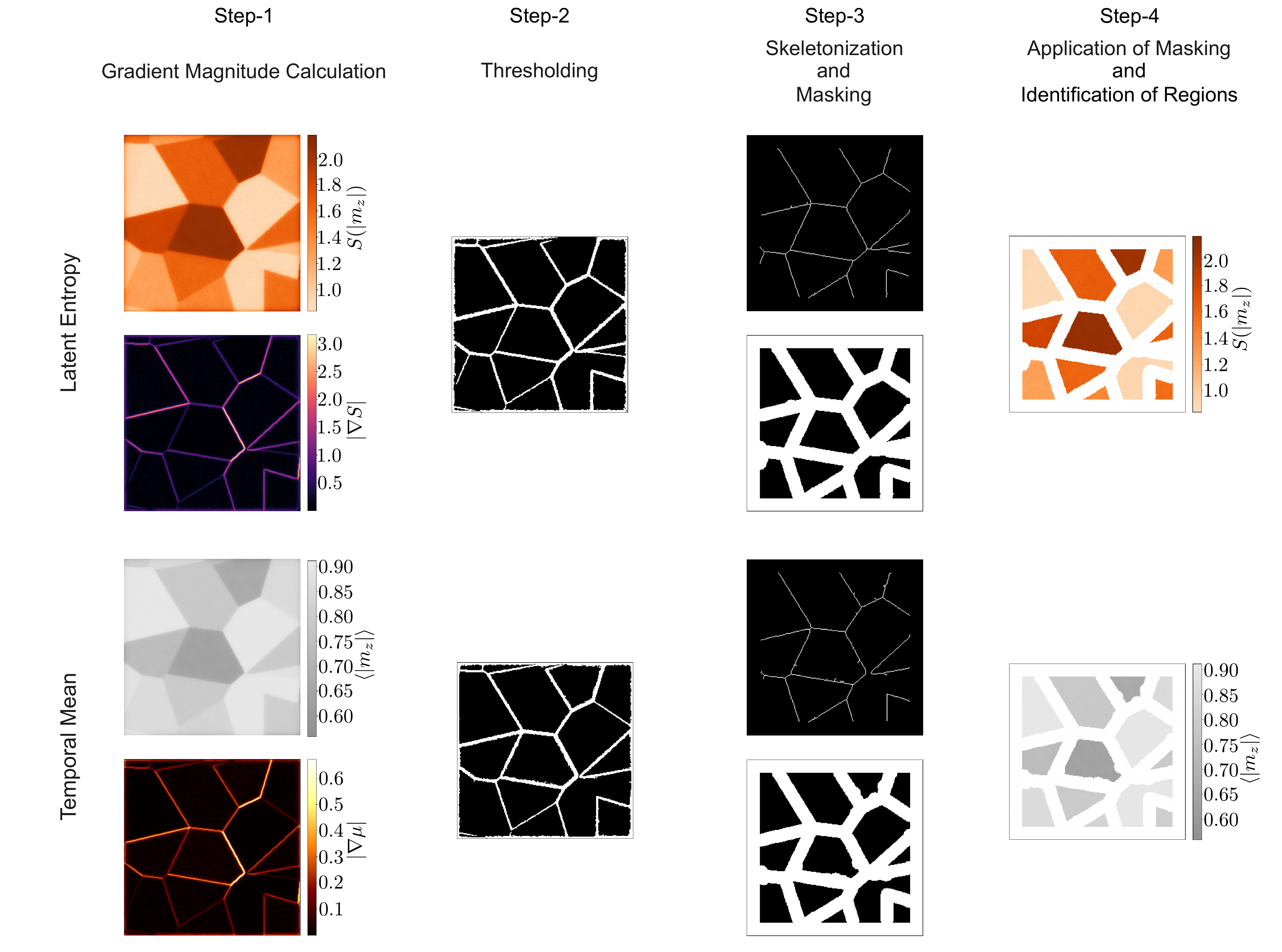} 
    \caption{Four-step procedure used to detect grain boundaries and construct exclusion masks before parameter inference. The top and bottom rows correspond to the latent entropy and temporal mean maps, respectively. Step~1 shows each spatial observable together with the magnitude of its spatial gradient. Step~2 shows the binary boundary masks obtained by hysteresis thresholding. Step~3 shows the skeletonized boundary centerlines and the corresponding dilated exclusion masks. Step~4 shows the masked observable maps and the remaining grain interiors used for pixel-wise parameter inversion. The white regions in Steps~2--4 denote the detected boundary and exclusion masks and are shown only for visualization; they do not represent numerical values of the corresponding descriptors.}
    \label{fig:FigD1_masking}
\end{figure*}

To localize the boundaries, the following four steps, shown in Fig.~\ref{fig:FigD1_masking}, were followed.

\noindent\textbf{Step 1: Gradient Magnitude Calculation}
Starting from the two spatial maps, $S (x, y)$ and $\mu (x, y)$, we compute the magnitude of their spatial gradients $|\nabla S(x,y)|$ and $|\nabla \mu(x,y)|$. At each interface, the gradient magnitude depends on the contrast in the corresponding observable between the two neighboring grains, with larger differences producing stronger boundary signals.

\noindent\textbf{Step 2: Thresholding}
The gradient magnitude maps are converted into binary boundary masks using hysteresis thresholding~\cite{Canny1986} with lower and upper thresholds, $\Delta_{\mathrm{low}}$ and $\Delta_{\mathrm{high}}$, as implemented in \texttt{scikit-image}~\cite{vanderWalt2014}. Pixels with values above $\Delta_{\mathrm{high}}$ are classified as confirmed boundary pixels, whereas those below $\Delta_{\mathrm{low}}$ are discarded as background noise. Pixels with intermediate values are retained only when they have at least a single neighbor which is a boundary pixel. This preserves weak but connected boundary segments while suppressing isolated fluctuations within the grain interiors. We use $(\Delta_{\mathrm{low}},\Delta_{\mathrm{high}})=(0.28,0.85)$ for $|\nabla S|$ and $(0.04,0.18)$ for $|\nabla\mu|$. The resulting masks are shown in Step~2 of Fig.~\ref{fig:FigD1_masking}.

\noindent\textbf{Step 3: Skeletonization and Masking}
First, an outer 19~nm frame is removed so that the sample edges are not misidentified as grain boundaries.
The thresholded boundary masks generally have non-uniform thickness because their width depends on the local gradient profile. We therefore reduce each mask to a one-pixel-wide boundary centerline using \texttt{skimage.morphology.skeletonize}~\cite{Zhang1984,vanderWalt2014}, which preserves the connectivity of the detected interfaces. The resulting centerlines are then expanded using \texttt{skimage.morphology.dilation} with a disk-shaped structuring element~\cite{vanderWalt2014}. With the $1$~nm cell size used here, dilation by $9$ pixels removes $9$~nm on each side of the one-pixel boundary skeleton, producing a total exclusion width of $19$~nm. This width is chosen to be comparable to the largest exchange length in the investigated parameter range, thereby reducing the contribution of interfacial exchange-coupled pixels while retaining sufficiently large grain interiors for parameter inference. The resulting skeletons and exclusion masks are shown in Step~3 of Fig.~\ref{fig:FigD1_masking}.
The boundary localization is limited by the contrast between neighboring grains. When adjacent regions have sufficiently similar values of $S$ or $\mu$, the corresponding gradient may remain below the hysteresis thresholds, causing the interface to go undetected and the two regions to be treated as a single connected region. In addition, dilation of the detected boundaries can fully mask narrow or small grain interiors, causing some regions to be excluded from the subsequent inversion or merged with neighboring connected regions.

\noindent\textbf{Step 4: Application of Masking and Identification of Regions}
The exclusion masks derived from Step~3 are applied separately to the corresponding latent entropy and temporal mean maps. 
Because the two descriptors produce boundary signals with different spatial continuity and noise levels, the resulting exclusion masks differ slightly. The remaining unmasked pixels define the grain interiors used for the subsequent pixel-wise parameter inversion, as shown in Step~4 of Fig.~\ref{fig:FigD1_masking}.

\subsection{Prediction in Heterogeneous Sample}
\begin{figure*}[htbp]
    \includegraphics[width=1.0\linewidth]{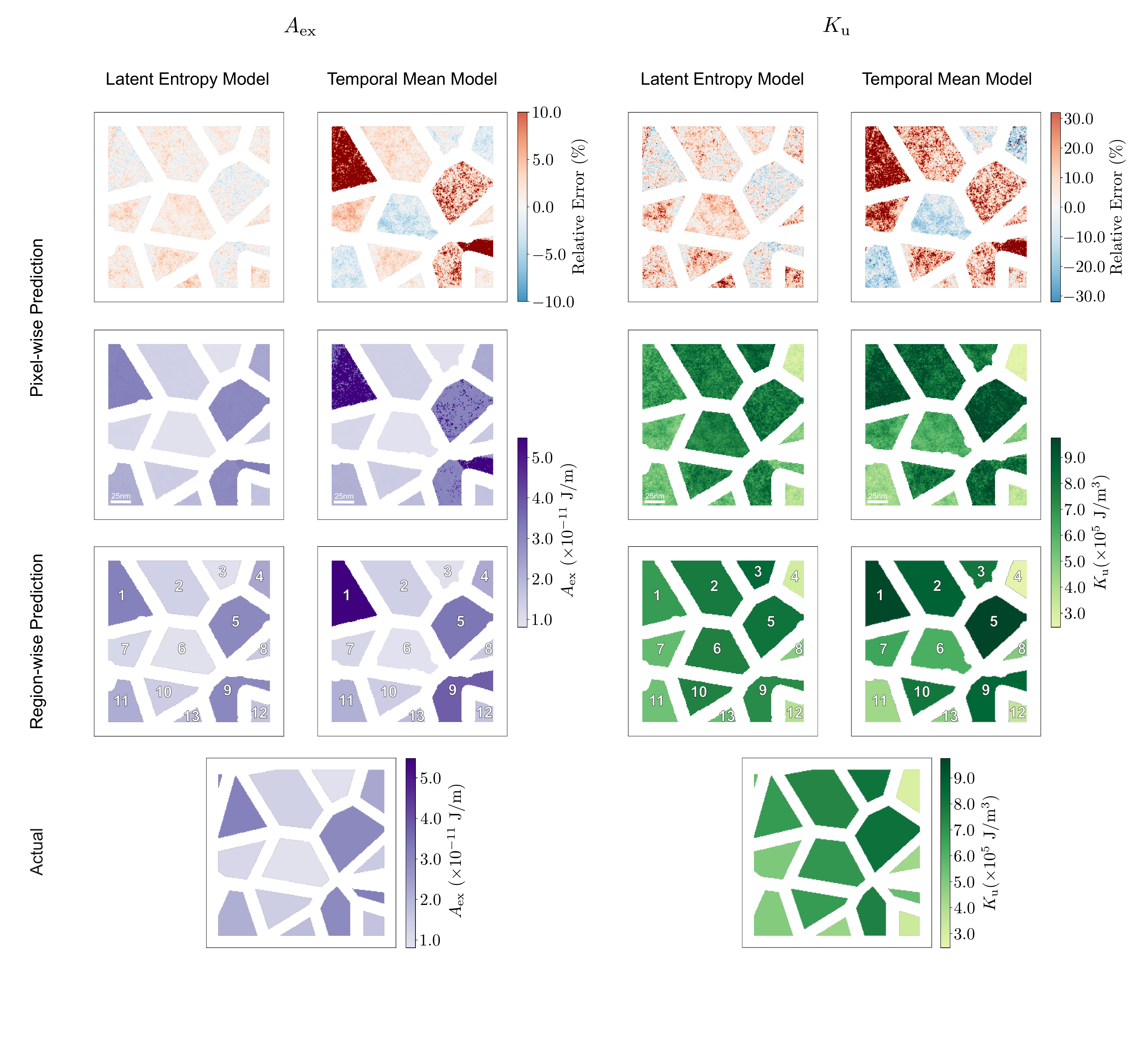} 
    \caption{Region-wise parameter predictions in the Voronoi-tessellated test sample. The left and right halves show the prediction of $A_{\rm ex}$ and $K_{\rm u}$, respectively, using the $S$- and $\mu$-models. The top row presents the pixel-wise relative prediction errors, the middle row the corresponding pixel-wise predicted parameter maps, and the bottom row the region-averaged predictions obtained from the unmasked pixels within each detected grain interior. The numbered regions correspond to the region-wise statistics reported in Table~\ref{tab:VoronoiRegion_error_compact}. White areas denote the excluded boundary and edge regions.}
    \label{fig:FigD2}
\end{figure*}

\begin{figure*}[ht]
    \includegraphics[width=0.9\linewidth]{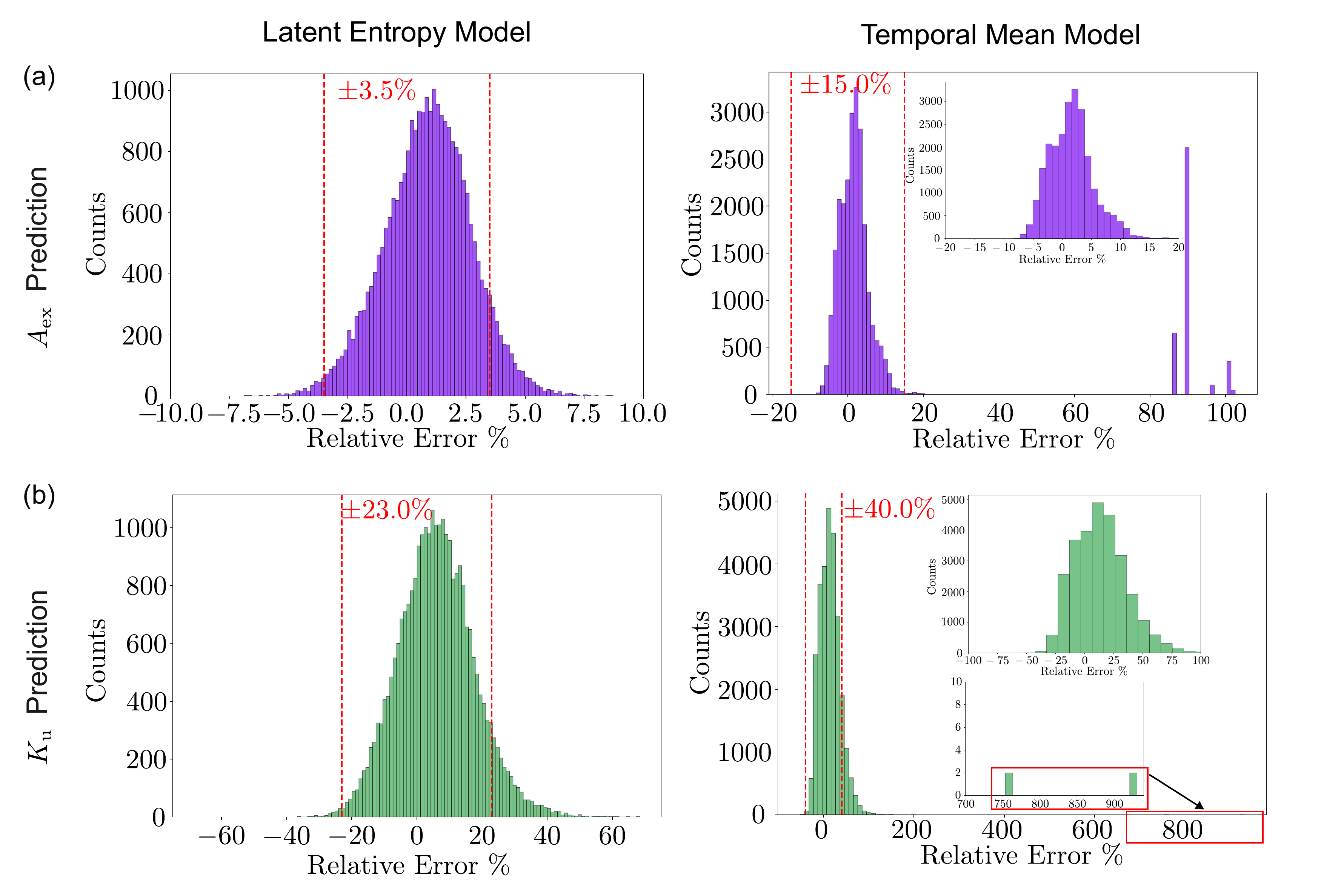} 
    \caption{
    Distributions of the relative reconstruction errors across the unmasked pixels of the Voronoi-tessellated sample shown in Figure~\ref{fig:Fig4}. The top and bottom rows correspond to the reconstruction of $A_{\rm ex}$ and $K_{\rm u}$, respectively, while the left and right columns show results from the $S$- and $\mu$-models. The dashed vertical lines indicate intervals containing approximately $90\%$ of the reconstructed values: $\pm3.5\%$ and $\pm15\%$ for $A_{\rm ex}$, and $\pm23\%$ and $\pm40\%$ for $K_{\rm u}$, from the $S$- and $\mu$-models, respectively. Insets in the $\mu$-model panels show magnified views of the central parts of the distributions, while the additional inset in the lower-right panel highlights the rare large-error outliers.}
    \label{fig:FigD3}
\end{figure*}

Following the masking procedure described in the previous part, the inference scheme is applied independently to each unmasked pixel of the $S$ and $\mu$ maps. 
Fig.~\ref{fig:FigD2} presents the resulting pixel-wise predictions, which are subsequently averaged within each detected region to obtain the region-wise results shown in Fig.~\ref{fig:Fig4} of the main text. 
The region-wise errors are reported in Table~\ref{tab:VoronoiRegion_error_compact}, while the full pixel-wise error distributions are shown in Fig.~\ref{fig:FigD3}. When a boundary is not resolved by the masking procedure, neighboring Voronoi grains are treated as a single connected region; the corresponding prediction values are therefore evaluated over the pooled pixels. Such merged regions are identified in Table~\ref{tab:VoronoiRegion_error_compact}.

For the prediction of $A_{\rm ex}$, the $S$-model produces comparatively uniform pixel-wise predictions within the grain interiors. Approximately $90\%$ of the pixel-wise errors lie within $\pm3.5\%$, as shown in Fig.~\ref{fig:FigD3}(a). The $\mu$-model exhibits a substantially broader distribution, with approximately $90\%$ of the errors lying within $\pm15\%$ and pronounced outliers in grains with large $A_{\rm ex}$. In particular, the region-averaged $A_{\rm ex}$ errors for regions~1, 5, and 9 reach approximately $75.2\%$, $12.6\%$, and $25.1\%$, respectively. The prediction of $K_{\rm u}$ is less accurate for both observables. For the $S$-model, approximately $90\%$ of the pixel-wise errors lie within $\pm23\%$, and the largest region-averaged absolute errors occur in regions~12 and 13, with values of approximately $15.4\%$ and $20.9\%$, respectively. The $\mu$-model yields an even broader and more strongly tailed distribution, with approximately $90\%$ of the errors lying within $\pm40\%$.

\begin{table*}[h]
\centering
\small
\setlength{\tabcolsep}{3pt}
\renewcommand{\arraystretch}{0.95}
\resizebox{\textwidth}{!}{%
\begin{tabular}{ccc cc cc}
\hline
\multirowcell{2}{\\\textbf{Region}} &
\multirowcell{2}{$\boldsymbol{\ell_{ex}}$\\\textbf{(nm)}} &
\multirowcell{2}{\textbf{Reference values}\\
\textbf{$(K_u,A_{ex})$}\\
\textbf{[$10^{5}$ J/m$^3$, $10^{-11}$ J/m]}} &
\multicolumn{2}{c}{\textbf{Avg. error ($S$-model)}} &
\multicolumn{2}{c}{\textbf{Avg. error ($\mu$-model)}} \\
\cline{4-7}
&
&
&
\textbf{\makecell{Rel.\\(\%)}} &
\textbf{\makecell{Abs. rel.\\(\%)}} &
\textbf{\makecell{Rel.\\(\%)}} &
\textbf{\makecell{Abs. rel.\\(\%)}} \\
\hline
 1$^{*}$ & 6.86 & (6.689, 3.151) & $(-1.27$, $-0.29)$ & $( 9.44$, $ 1.59)$ & $(42.76$, $73.65)$ & $(42.77$, $73.66)$ \\
 2       & 4.24 & (7.433, 1.338) & $( 3.52$, $ 0.56)$ & $( 7.84$, $ 1.32)$ & $(17.71$, $ 2.13)$ & $(18.20$, $ 2.20)$ \\
 3       & 3.27 & (8.111, 0.866) & $( 2.84$, $ 0.46)$ & $( 6.57$, $ 1.13)$ & $(-2.29$, $-0.59)$ & $( 7.75$, $ 1.54)$ \\
 4       & 8.85 & (2.900, 2.273) & $( 5.59$, $ 0.69)$ & $(10.35$, $ 1.35)$ & $(-15.06$, $-2.02)$ & $(16.05$, $ 2.15)$ \\
 5       & 6.04 & (8.218, 2.994) & $(-3.80$, $-0.73)$ & $( 8.58$, $ 1.54)$ & $(18.68$, $12.63)$ & $(19.48$, $12.88)$ \\
 6       & 3.46 & (6.920, 0.829) & $( 7.43$, $ 1.20)$ & $( 8.44$, $ 1.37)$ & $(-10.88$, $-2.62)$ & $(11.48$, $ 2.74)$ \\
 7       & 4.80 & (5.066, 1.165) & $( 8.23$, $ 1.20)$ & $(11.15$, $ 1.67)$ & $(26.95$, $ 3.81)$ & $(26.99$, $ 3.81)$ \\
 8       & 5.88 & (4.312, 1.490) & $(10.45$, $ 1.47)$ & $(12.16$, $ 1.73)$ & $(14.98$, $ 2.00)$ & $(15.47$, $ 2.07)$ \\
 9$^{*}$ & 6.49 & (7.167, 3.022) & $(-2.42$, $-0.47)$ & $( 8.84$, $ 1.52)$ & $(23.32$, $25.41)$ & $(24.33$, $25.70)$ \\
10       & 4.56 & (6.776, 1.412) & $( 8.51$, $ 1.34)$ & $(11.07$, $ 1.78)$ & $(17.09$, $ 2.16)$ & $(17.80$, $ 2.27)$ \\
11$^{\dagger}$ & 6.62 & (4.920, 2.153) & $( 6.44$, $ 0.93)$ & $(10.07$, $ 1.50)$ & $(-12.32$, $-2.20)$ & $(13.12$, $ 2.34)$ \\
12       & 7.14 & (3.266, 1.667) & $(14.52$, $ 1.88)$ & $(15.40$, $ 2.01)$ & $( 9.29$, $ 1.10)$ & $(11.39$, $ 1.36)$ \\
13       & 6.37 & (4.498, 1.823) & $(20.43$, $ 2.86)$ & $(20.86$, $ 2.93)$ & $( 5.45$, $ 0.77)$ & $( 9.37$, $ 1.36)$ \\
\hline
\end{tabular}%
}
\caption{Voronoi-tessellated sample region-wise relative error statistics for $S$- and $\mu$-model-based predictions.\\
{\small
All ordered pairs in the reference-value and error columns are given as
$(K_{\rm u},A_{\rm ex})$. After excluding edge-touching regions, 17 interior Voronoi regions remain.
For the region-wise error analysis, the descriptor-dependent masks reduce these
to 13 effective evaluated regions, listed in the first column. Regions 14--17
denote interior regions that are not evaluated separately because they are
either merged with neighboring evaluated regions or removed by the mask.\\[2pt]
$^{*}$Merged region in both $S$-model and $\mu$-model inversions;
$(K_{\rm u}, A_{\rm ex})$ values are the pooled-pixel weighted averages.
$S$-model: region~1\,$\leftarrow$\,[1,\,14]; region~9\,$\leftarrow$\,[9,\,15].
$\mu$-model: region~1\,$\leftarrow$\,[1,\,14];
region~9\,$\leftarrow$\,[9,\,15,\,16].\\[2pt]
$^{\dagger}$Merged in $\mu$-model inversion only:
region~11\,$\leftarrow$\,[11,\,17]; in the $S$-model, region~17 is masked out.
}}
\label{tab:VoronoiRegion_error_compact}
\end{table*}

\bibliographystyle{apsrev4-2}
\bibliography{bibliography}
\end{document}